# Magnetic Field and Frequency Dependent AC Susceptibility of High-$T_\text{c}$ YBCO Single Crystal

*M. Rakibul Hasan Sarkar, S. H. Naqib\**
*Department of Physics, University of Rajshahi, Rajshahi-6205, Bangladesh*
*\*Corresponding author email: salehnaqib@yahoo.com*

**Abstract**

The temperature dependence of AC susceptibility (ACS) has been measured for a very high-quality plate-like slightly overdoped YBCO single crystal for different frequencies and AC magnetic field amplitudes. Frequency dependence of the ACS is weak irrespective of the magnetic field orientation but significant effects of field orientation with respect to the $CuO_2$ planes and field magnitude on real and imaginary components of fundamental ACS were observed. We have found that the magnitude of the ratio of the loss peak to the full diamagnetic signal is slightly larger in perpendicular field than that in the parallel field arrangement. The height of the loss peak saturates as full penetration of magnetic field is achieved. This saturation is achieved at a lower field amplitude when H∥c. The peak temperature, $T_p$, in $\chi''$ shifts to lower temperatures with increasing magnetic field amplitude for both H∥c and H∥ab. The value of $T_p$ depends on the orientation of the magnetic field with respect to the crystallographic axes, illustrating the anisotropy in the magnetic flux dynamics. The superconducting onset transition temperature obtained from the diamagnetic shielding component of the ACS, on the other hand, does not depend on the orientation of the magnetic field. The superconducting transition width increases weakly with increasing magnetic field, the rate of increment is relatively more prominent for H∥c case. The Cole-Cole plot [$\chi''(\chi')$] shows qualitatively and quantitatively identical features for H∥c and H∥ab, independent of the orientation of the magnetic field with respect to the sample geometry and shielding current paths. The general features of $\chi''(\chi')$ implies that, there is no flux creep for the range of frequencies and AC fields employed in this investigation. The maximum value of the loss peak and its position with respect to $\chi'$ in the Cole-Cole plot are largely consistent with the Bean critical state model. Slightly increased peak value in comparison to the predicted peak value within the Bean critical state model is probably due to a weak field dependence of $J_c$. The results obtained here are compared with various theoretical models and experimental findings. Prominent differences are noted and discussed in details in this study.

## 1. Introduction

In recent years, among all cuprate high-$T_c$ superconductors, the double $CuO_2$ layer $YBa_2Cu_3O_{7-\delta}$ (YBCO) stands as the most extensively studied compound. YBCO is also considered as one of the most promising systems for high critical current density and high magnetic field applications [1, 2]. Extensive research has been carried out on temperature, field and frequency dependent AC magnetic susceptibility of high-$T_c$ $YBa_2Cu_3O_{7-\delta}$ superconductors to investigate the critical current density ($J_c$), irreversibility magnetic field and vortex dynamics of this compound [3-6].

AC susceptibility (ACS) is among the most widely employed technique for understanding a number of basic superconducting state properties. Even though the diamagnetic transition of superconductors is well understood, the interpretation of the complex ACS data in the mixed state remained largely empirical, more so for high-$T_c$ cuprates which are characterized by high level of structural and electronic anisotropies. Among the proposed schemes, the critical state model (CSM) [2, 7, 8] is the most widely employed one for explaining the temperature-dependent features of ACS. The Anderson-Kim model [8] has been used extensively for comparison with experimental results in terms of the temperature and field dependence of the critical current density $J_c$(H, T) [9-13]. However, it is to be noted that although the CSM has explained a broad range of experimental findings with significant success, as a static hysteretic model, CSM is incapable of explaining the frequency-dependent ACS of high-$T_c$ cuprates. In the presence of flux creep and thermally assisted flux flow (TAFF) regime, frequency (f) of the AC field becomes a major parameter controlling the temperature and field dependent characteristics of the ACS signal. The structural and electronic anisotropies add further complications to the description of vortex dynamics for different magnetic field orientations with respect to the crystallographic axes. All these resulted in a significant interest in the study of ACS techniques to understand the loss mechanisms in high-$T_c$ cuprates related to TAFF, giant flux creep, vortex glass phase, temperature evolution of the irreversibility field and possible melting transition of flux line lattice [14-16]. Besides, the richness of the magnetic phase diagram has been a matter of fundamental scientific interest [14-16]. A thorough understanding of the vortex dynamics of high-$T_c$ cuprates is essential to minimize loss and maximize the critical current density in these compounds. In this context, understanding the temperature, field and frequency dependent real and imaginary parts of AC magnetic susceptibility is of significant importance for potential applications and extraction of the critical current density of high-$T_c$ cuprate superconductors.

The measurement of complex AC magnetic susceptibility $\chi$(T, H, f) = $\chi'$(T, H, f) + $i\chi''$(T, H, f) as a function of temperature, field and frequency provide information related to a number of physical properties of high-$T_c$ cuprates including the critical current density which is vital as far as the industrial applications are concerned. The critical current density can be determined from the temperature and field variation of AC susceptibility data using the modified critical state models (MCSMs) [3]. The real part of AC susceptibility discloses the dispersive magnetic response which is a measurement of magnetic shielding capability of the superconductor generated by supercurrent. The magnetic energy stored in the sample is proportional to χ'. The imaginary part corresponds to the ac losses that represent the amount of AC magnetic energy converted into heat in the sample and the energy loss per cycle is proportional to χ". The temperature, field and frequency dependent profiles of χ' and χ" plots are strongly influenced by structural/electronic anisotropy of the compound under consideration, fabrication technique, heat treatment, microstructure, orientation of applied field, and the measuring conditions [3-6]. The ACS modeled with the critical state approach is quasistatic in nature. This implies that the ACS signal depends on the magnitude of the magnetic field only but not on the frequency. This is true even when $J_c = J_c(H)$. In contrast to nonlinear quasistatic ACS, the linear ACS with frequency dependent resistive state, depends only on the frequency of the AC field but not on the magnitude of the magnetic field. Such contrasting behaviors provide one with a robust framework to analyze the ACS data and find out the nature of the flux dynamics.

Variety of approaches have been adopted by researchers to analyze the temperature, field, and frequency dependent ACS signal [17-21]. Herzog et al. [18] used the following equations to express the temperature and field dependent real and imaginary components of complex AC susceptibility for thin superconducting disk under perpendicular magnetic field configuration:

$$\chi'(T) = -\frac{1}{x(T)}\tanh x(T) \tag{1}$$

$$\chi''(T) = -\frac{1}{x(T)}\tanh x(T) + \frac{2}{x(T)}\tanh \frac{x(T)}{2} \tag{2}$$

where, $x(T) = \frac{H_{ac}}{H_d(T)}$ and $H_d(T) = \frac{J_c(T)\, d}{2} = J_{0c}(1 - \frac{T}{T_c})^\beta \frac{d}{2}$

Here, $H_{ac} = H_{rms}$ is the root-mean-squared (rms) value of the magnetic field, d is the thickness of the sample, $J_c$ is the critical current density, $J_{0c}$ is the zero temperature critical current density, and β is a material dependent power exponent closely linked with the flux pinning mechanism within a sample.

Clem and Sanchez [19] deduced and used the following equations as theoretical expressions to represent χ' and χ" for a thin disk of radius R and thickness d in a perpendicular AC magnetic field:

$$\chi' = \frac{2\chi_0}{\pi}\int_0^\pi (1 - \cos\theta)S[(x/2)(1-\cos\theta)]\cos\theta\, d\theta \tag{3}$$

$$\chi'' = \frac{2\chi_0}{\pi} \int_0^\pi \{-S(x) + (1 - \cos\theta)S[(x/2)(1 - \cos\theta)]\}\sin\theta \, d\theta \qquad (4)$$

where, $x = \frac{H_{ac}}{H_d}$, $H_d = \frac{J_c d}{2}$, $\chi_0 = \frac{8R}{3\pi}$, and

$$S(x) = \frac{1}{2x}\left[\cos^{-1}\left(\frac{1}{\cosh(x)}\right) + \frac{\sinh(x)}{\cosh^2(x)}\right]$$

Assuming Bean's CSM where the critical current density is independent of the magnetic field, Brandt [20, 21] found the following relations to express χ' and χ":

$$\chi'' \approx 0.8|\chi'|^{2/3}(1 - |\chi'|)^{1.19} \quad \text{for all values of } 0 \leq \chi' \leq -1 \qquad (5)$$

It has been found that in the low temperature limit, the relation between χ' and χ" for the thin film geometry is given by,

$$\chi'' = \alpha(1 + \chi')$$

with $\alpha = 0.68$ when the AC magnetic field is perpendicular to the surface of the film [17, 20, 21].

Besides, a number of models have been put forward with different forms of $J_c$(H) within the MCSM to model the complex ACS explicitly as functions of temperature, magnetic field and frequency [22-24] with varying degrees of success.

As far as the situation regarding experimental results on the complex ACS is concerned, there is significant ambiguity. A large number of ACS studies have been carried out [17, 19-21, 25-28]. Most of these studies have been on thin films and sintered polycrystalline high-$T_c$ cuprates belonging to different classes. Comprehensive study of experimental ACS for single crystals is very rare. Moreover, the state of doping has been rarely taken into account rigorously. It is known that the normal and superconducting state properties of cuprates depend strongly on the hole content in the $CuO_2$ planes [29-31]. Therefore, study of experimental ACS of high-quality single crystals with well-defined doping state is important to understand the intrinsic flux dynamics of high-$T_c$ cuprates. Against this backdrop, we have studied the temperature, field and frequency dependent real and imaginary parts of ACS of slightly overdoped high-quality $YBa_2Cu_3O_{7-\delta}$ single crystal for both H∥c and H∥ab magnetic field configurations. The characteristic features of the ACS have been investigated in detail.

Rest of the paper has been organized as follows. Information about the experimental single crystal is given and representative ACS results are displayed in Section 2. The ACS results are analyzed in Section 3. Important aspects of the results and analyses are discussed and summarized in Section 4.

## 2. Experimental sample and results of ACS measurement

A very high-quality slightly overdoped plate-like YBCO single crystal was used. The crystal was grown from flux in a $Y_2O_3$ crucible (with 1% BaO as binder) [32]. Further details regarding the sample characterization and state of oxygenation can be found in Refs. [33, 34]. The AC magnetic susceptibility of the single crystal was measured as a function temperature, ranging from 95 K to 85 K, i.e, from just above to below the superconducting transition temperature. The experiments were done at 1, 3, 10 and 13 Oe of AC magnetic fields (RMS values) with frequencies of 33.33 Hz, 333.33 Hz, 666.66 Hz and 1000 Hz at each AC magnetic field. The AC susceptibility of the sample was measured using a commercial Lake Shore AC susceptometer where the induced voltage was the representation of AC susceptibility. The in-phase voltage gave real part ($\chi'$) of ACS and the out-of-phase voltage gave imaginary part ($\chi''$) of ACS. The dimension of the rectangular single crystal was as follows; 1.82 mm x 1.59 mm with a thickness of 0.4 mm. Measurements of room temperature thermopower and superconducting $T_c$ indicate that the single crystal under study is slightly overdoped with a hole content ~0.18 [35, 36] with an oxygen deficiency $\delta$ ~0.03 in $YBa_2Cu_3O_{7-\delta}$. High level of oxygen loading was achieved by annealing the single crystal in flowing oxygen for ~10 days at a temperature of 420°C. The AC susceptibility has been measured for both H∥c and H∥ab configurations. Some representative field (RMS value) and frequency normalized ACS plots showing raw $\chi'(T)$ and $\chi''(T)$ are displayed below.

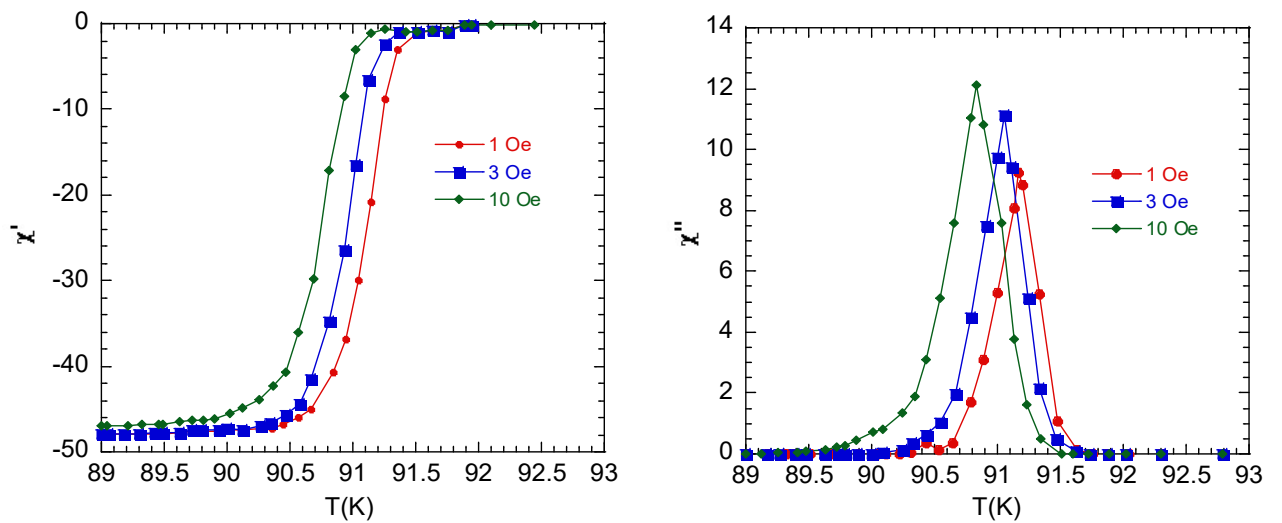

Fig. 1(a): Field normalized $\chi'(T)$ and $\chi''(T)$ as a function of applied AC (RMS) fields at 33.33 Hz. The magnetic field has been applied along the c-axis direction (H∥c).

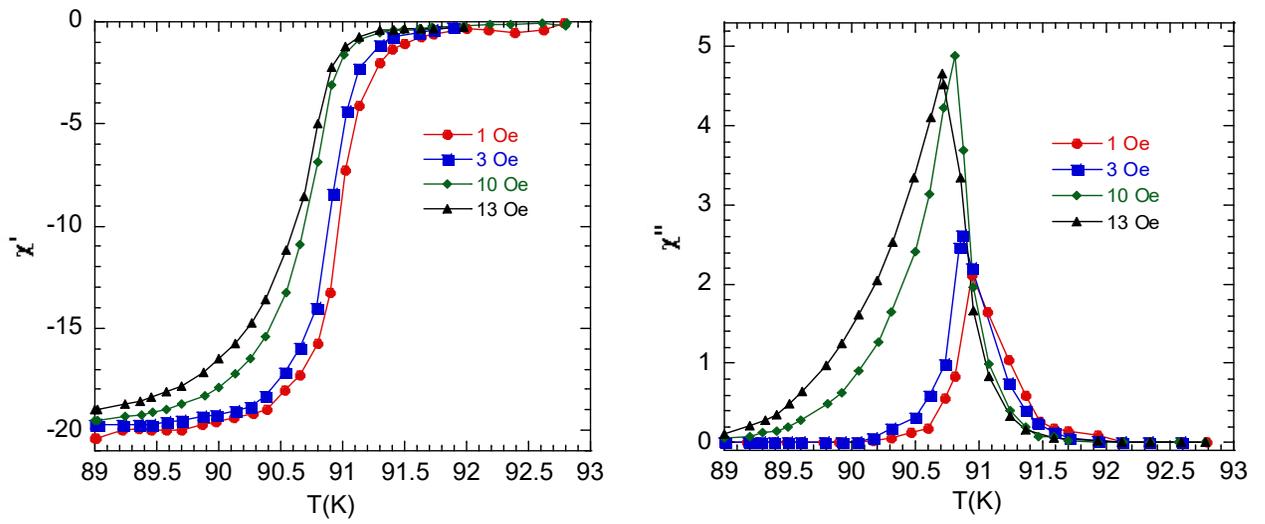

Fig. 1(b): Field normalized χ'(T) and χ"(T) as a function of applied AC (RMS) fields at 33.33 Hz. The magnetic field has been applied in the ab-plane (H||ab).

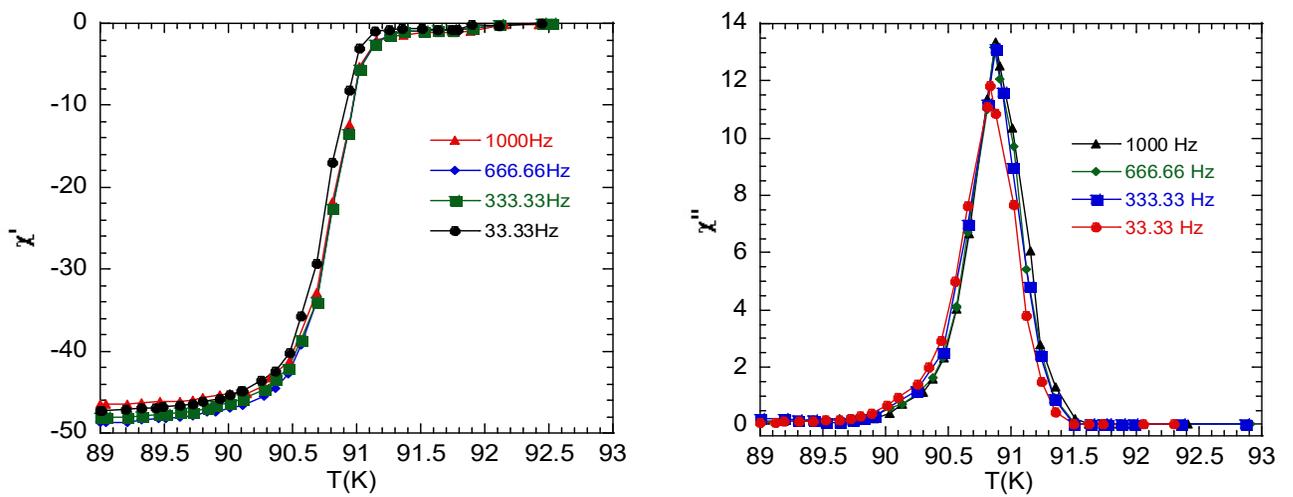

Fig. 1(c): Frequency normalized χ'(T) and χ"(T) as a function of frequencies at magnetic field 10 Oe applied along the c-axis direction (H||c).

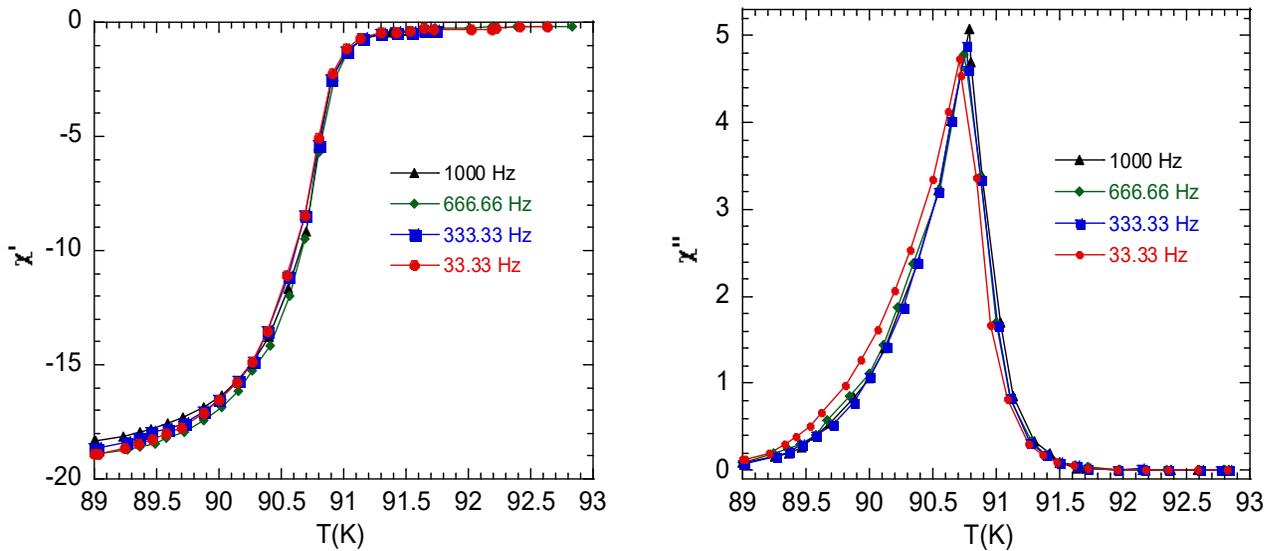

Fig. 1(d) : Frequency normalized χ'(T) and χ"(T) as a function of frequencies at magnetic field 13 Oe applied parallel to the ab-plane.

Mere inspection of Figs. 1 reveals several interesting features. Compared to prior experimental results [17, 19-21, 25-28], the temperature width of the superconducting transition is appreciably narrower at all magnetic fields and frequencies for our single crystal. Transition width appears relatively larger for the H∥ab configuration compared to that for the H∥c configuration. This is because the field induced tailing effects in χ'(T) and χ"(T) are more prominent in the case of H∥ab at low temperature. The peak value of χ"(T) depends strongly on the magnitude of the applied magnetic field at low field values and it saturates for applied fields of 10 Oe and above. The saturation of χ'(T) at low temperatures below $T_c$ implies that full diamagnetic shielding has been achieved at all values of the applied magnetic fields. Moreover, the field and frequency normalized ACS is almost 2.5 times larger in case of H∥c compared to that in the configuration H∥ab. There are two main possible factors: i) shielding supercurrent is larger when it flows entirely in the $CuO_2$ plane (H∥c configuration) and ii) there is significant effect of large demagnetization factor in this particular field configuration for thin superconducting samples [22, 37-39].

3. Analysis of the ACS data

As far as the ACS is concerned, it is well known that its real part is a measurement of the magnetic shielding capability of the superconductor generated by supercurrents and the imaginary part is related to losses. For high quality single crystals with dimensions much larger than the superconducting penetration depth λ, the absolute value of χ'(T) rises sharply at $T_c$ as the diamagnetic screening increases with the lowering of temperature. χ"(T) shows a peak below $T_c$. At even lower T, both χ'(T) and χ"(T) curves become flat, demonstrating complete diamagnetic shielding and low-loss states, respectively. The raw ACS data shown in Figs. 1 imply that full diamagnetic shielding is achieved at lower temperature limits of measurement. Therefore, the maximum values of χ'(T) has been normalized to -1.0 and also the χ"(T) values using the same proportion accordingly. We have shown representative normalized plots of χ'(T) and χ"(T) in Figs. 2 which illustrate the effect of magnetic field at fixed frequencies.

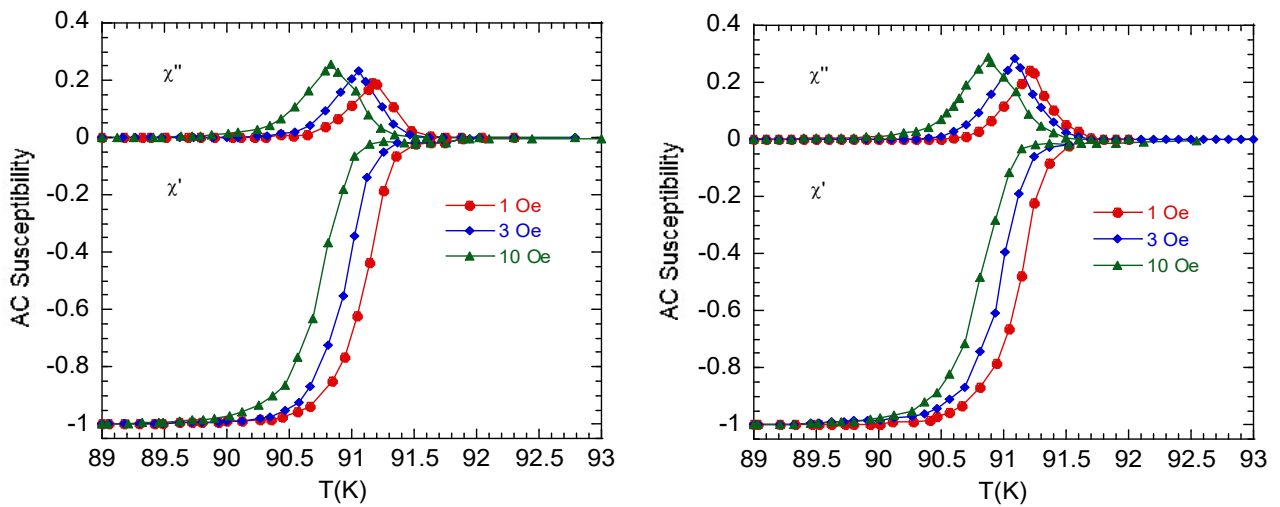

Fig. 2(a): Real (χ') and imaginary (χ") components of the complex ac-susceptibility as a function of temperature at 33.33 Hz and 1000 Hz frequencies (left and right panels) for YBCO single crystal at various AC (RMS) fields, 1 Oe, 3 Oe and 10 Oe. The magnetic field is applied parallel to c-axis of the crystal.

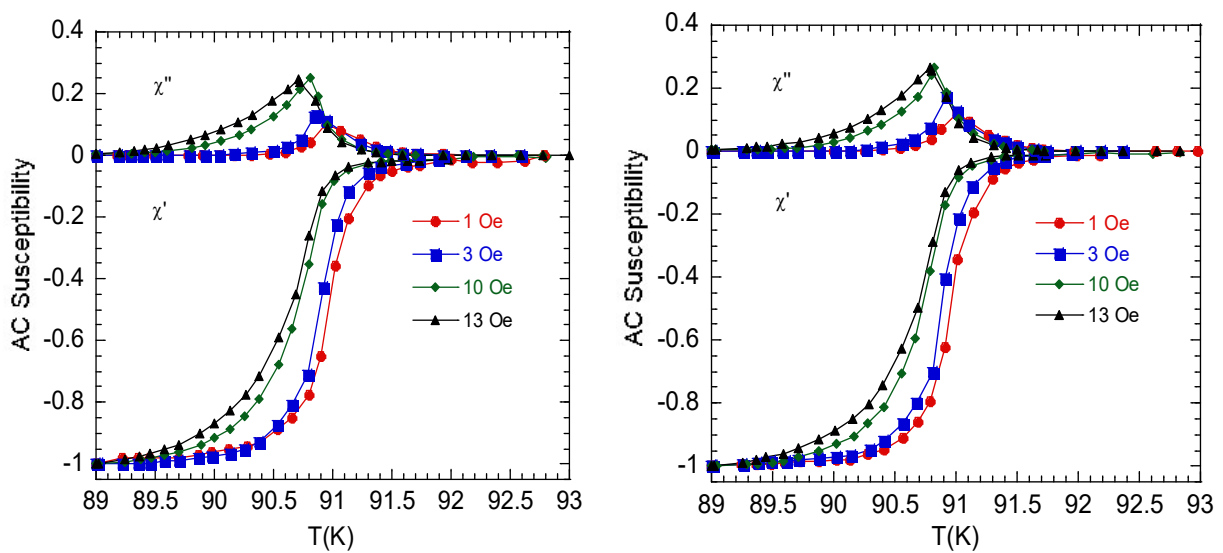

Fig. 2(b): Real (χ') and imaginary (χ") components of the complex ac-susceptibility as a function of temperature at 33.33 Hz and 1000 Hz frequencies (left and right panels) for YBCO single crystal at various AC (RMS) fields, 1 Oe, 3 Oe, 10 Oe and 13 Oe. The magnetic field is applied parallel to ab-plane of the crystal.

To illustrate the effect of frequency, we have plotted χ'(T) and χ"(T) data at fixed magnetic fields in Figs. 3.

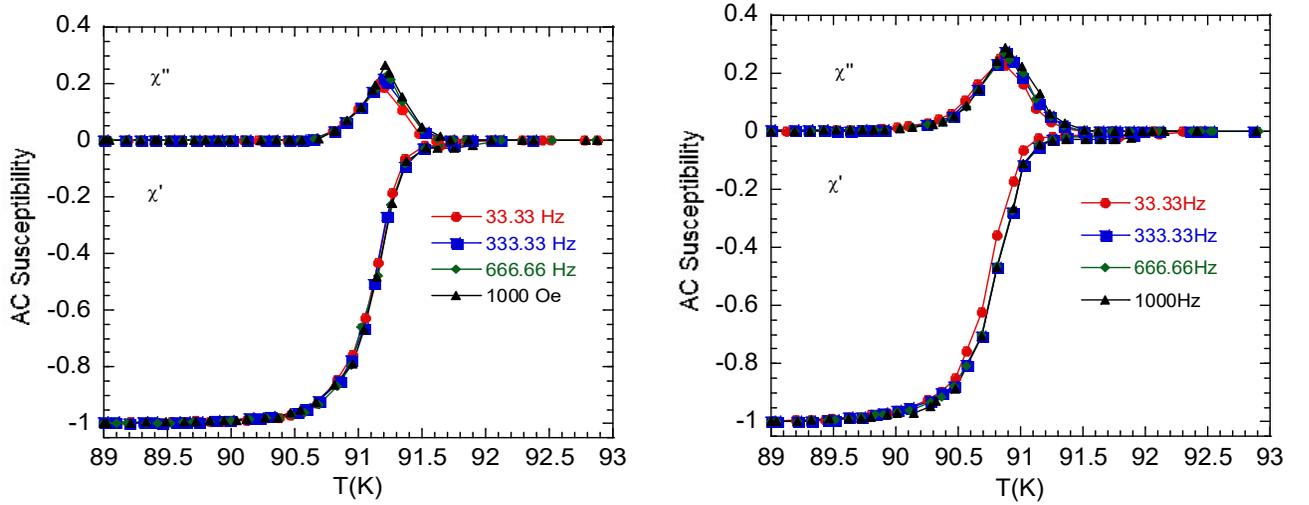

Fig. 3(a): Variation of χ'(T) and χ"(T) at 1 Oe and 10 Oe fields (left and right panels) for YBCO single crystal with different AC frequencies. The magnetic field is applied parallel to the c-axis of the crystal.

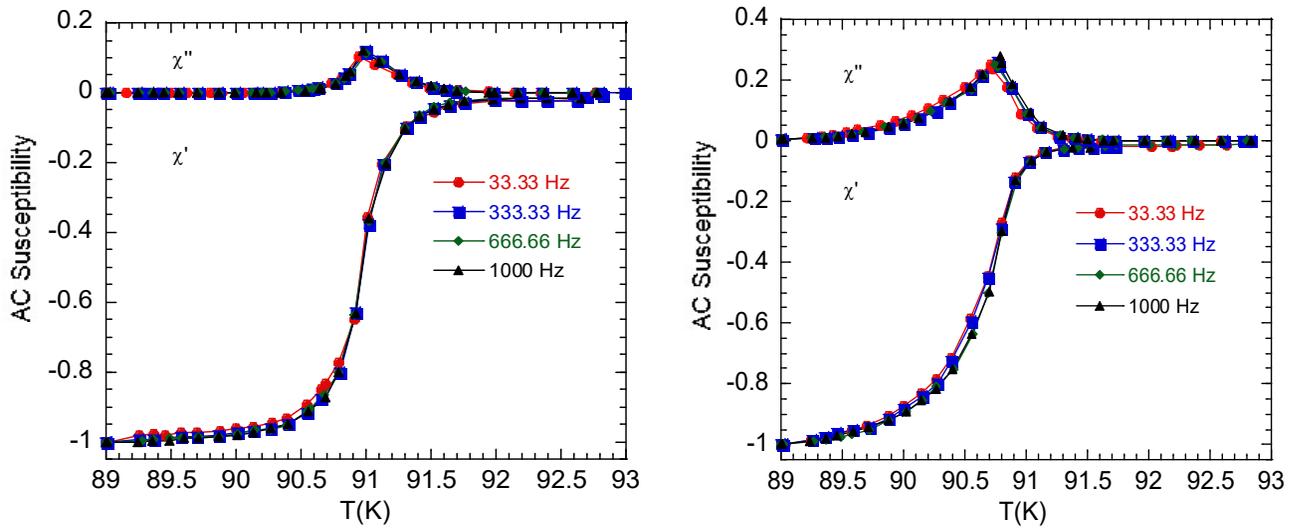

Fig. 3(b): Variation of χ'(T) and χ"(T) at 1 Oe and 13 Oe fields (left and right panels) for YBCO single crystal with different AC frequencies. The magnetic field is applied parallel to the ab-plane of the crystal.

The ACS profiles are characterized by several parameters, namely, the peak temperatures ($T_p$) in χ"(T), transition widths of χ'(T) and, the χ"(T) versus χ'(T) plot (known the Cole-Cole plot). We investigate each of these characteristic features below with different magnetic fields, frequencies and field orientations. The variations of $T_p$ in χ" with magnetic field for H||c and H||ab configurations are shown in Fig. 4. Since $T_p$ is insensitive to frequency, only the situation for f = 1000 Hz is illustrated in Fig. 4.

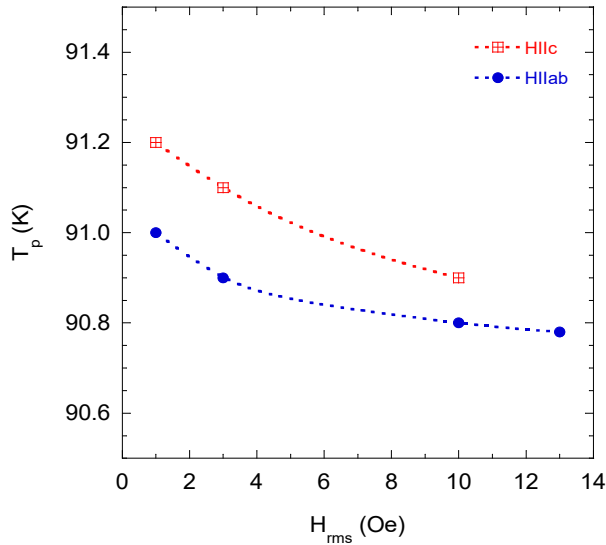

Fig. 4: Variation of $T_p$ in $\chi''$ (f = 1000 Hz) with magnetic field with H∥c and H∥ab configurations. The dashed lines are drawn as guides to the eye.

It is interesting to note that $T_p$ is higher for H∥c compared to that for H∥ab. This shows the anisotropy in flux dynamics in YBCO. The variation of $T_p$ with magnetic field can be qualitatively explained with the aid of the non-linear response CSM [3, 40-42]. In this formalism, ACS results from the hysteretic penetration of magnetic fluxoids. Hysteretic behavoir arises due to the presence of flux pinning centers inside a superconductor. Qualitatively, the peak in $\chi''(T)$ in the CSM originates when the magnetic field reaches the center of the sample. $T_p$ is field dependent in the CSM but it has no frequency dependence unless $J_c$ itself is frequency dependent, or there is a possibility of flux creep inside the superconductor [3, 40, 43]. The maximum magnitude of peak ratio, defined as the ratio of the peak value of $\chi''(T)$ to full diamagnetic susceptibility (of magnitude 1), lies within the range 0.22-0.25 [3, 40] in the CSM when $J_c$ is independent or weakly dependent on the magnetic field H. For the YBCO single crystal under consideration, no frequency dependence of $T_p$ was observed for $\chi''(T)$ (Figs. 3a and 3b), consistent with simple CSM. $T_p$ shifts towards lower temperature with increasing AC field amplitudes (Fig. 4), which also has a very satisfactory qualitative explanation within the CSM [3, 40]. As AC field amplitude increases a larger $J_c$ is needed to screen the applied magnetic field. $J_c$ increases with the lowering of temperature as superconducting (SC) condensate is enhanced [2, 44, 45], therefore, $T_p$ decreases with increasing magnetic field. The width of $\chi''(T)$ peak increases steadily with increasing magnetic field for both H∥c and H∥ab configurations.

Field dependent SC transition temperatures are also determined. The methodology is disclosed in Fig. 5(a). The diamagnetic onset temperature of intrinsic superconducting transition is about 91.70 K for our experimental YBCO sample in H∥c case for 1000 Hz and 1 Oe applied field. This is the temperature at which phase coherent Cooper pairs are formed.

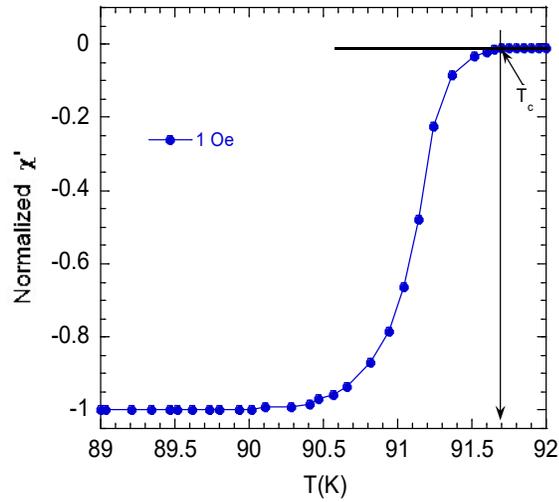

Fig. 5(a): Determination of superconducting transition temperature from the $\chi'(T)$ onset temperature of downward deviation from the normal state background ACS signal.

The superconducting onset is insensitive for the range of frequency used in this study. The field dependent $T_c$ values for both H∥c and H∥ab are shown in Fig. 5(b).

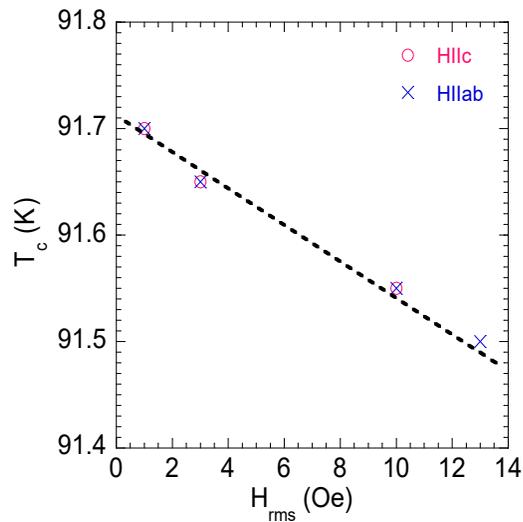

Fig. 5(b): Onset temperature for superconducting transition at different magnetic fields (f = 1000 Hz). A linear fit to the $T_c(H_{rms})$ data is shown.

The onset temperature for superconducting transition shows no dependency on the magnetic field orientation. Next, we have shown the peak temperature ($T_d$) in the first-derivative of $\chi'(T)$ ($d\chi'(T)/dT$) together with the full-width at half maximum (FWHM), defined as the width of the SC transition in Fig. 6. Both $T_d$ and FWHM are fairly insensitive to the frequency of the AC magnetic field.

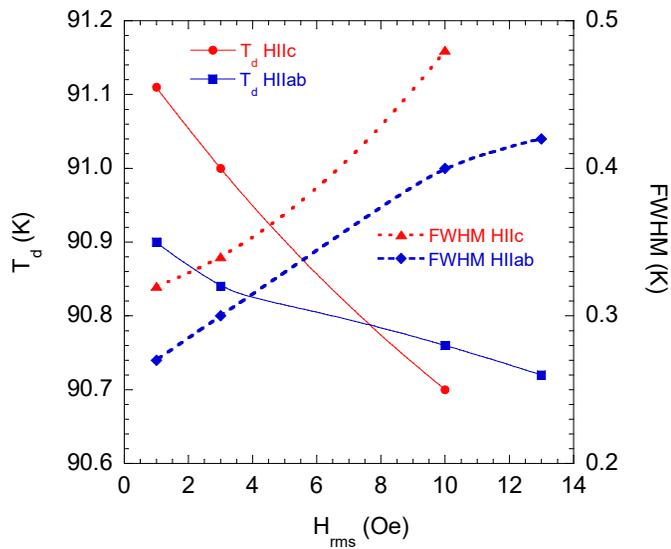

Fig. 6: $T_d$ and FWHM of $d\chi'(T)/dT$ curve for YBCO single crystal at different magnetic fields. The lines are drawn as guides to the eye. The data points are obtained at a AC frequency of 33.33 Hz.

It is interesting to note that $T_c$ is insensitive to field orientation (Fig. 5b) but $T_d$ depends on the field configuration with respect to the crystallographic axes. The shift in $T_d$ with magnetic field is much stronger for H∥c configuration compared to H∥ab case. The transition width is also larger for the H∥c configuration. This discloses the fact that the SC transition width is affected by the anisotropy in the vortex dynamics in the sample. It is curious to note that, there is a clear correspondence between the temperatures $T_p$ and $T_d$. Except for a small offset of ~0.10 K, these two temperatures are almost identical to each other. We illustrate this for H∥c in Fig. 7. The same qualitative behavior is observed when H∥ab.

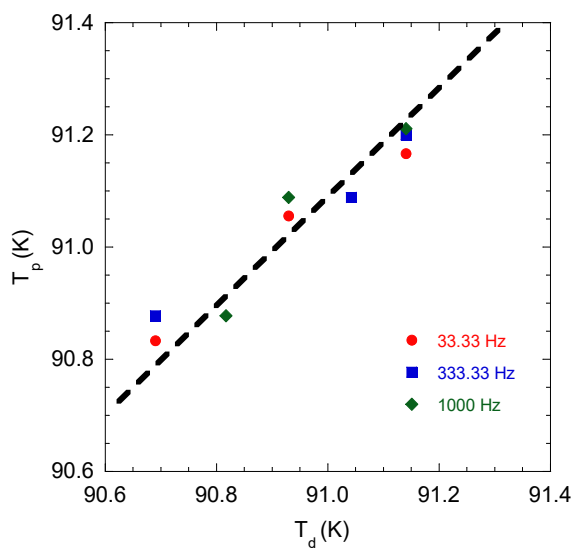

Fig. 7: $T_p$ versus $T_d$ of YBCO single crystal at different AC frequencies and magnetic fields for H∥c configuration. The thick dashed line displays the general trend.

In contrast to ACS results as a function of the field amplitude, frequency, or temperature, a parametric plot of $\chi''(T)$ versus $\chi'(T)$ contains only dimensionless quantities, and is therefore very useful for analyzing experimental data. Using this parametric representation, a comparison between experimental ACS data and theoretical ACS models can be made without invoking any particular temperature dependence of the critical current density inside the sample. It is important to realize that, in general, $J_c$ is a function of h (reduced magnetic field) defined as, $h = H_{rms}/H_p$ with $H_p = J_c d$; d being an effective dimensional parameter of the sample under consideration and $H_p$ is the field at which the magnetic flux reaches at the center of the sample. Therefore, when $\chi''(T)$ is plotted against $\chi'(T)$, the reduced magnetic field should be treated as a curve parameter lying within the range $0 \leq h < \infty$. Under these circumstances, the specific shape of the Cole-Cole plot is determined by the sample geometry, magnetic field configuration and the details of flux dynamics including the functional form of the field dependence of $J_c$. We have shown the representative Cole-Cole plots both with varying magnetic fields and frequencies in Figs. 8. A number of interesting features are seen in these Cole-Cole plots. The peak value in the $\chi''(\chi')$ plot depends on the magnitude of the magnetic field. At low fields (1 Oe in case of H∥c and 1 Oe and 3 Oe in case of H∥ab) the peak values are small which increase and saturate at higher fields. This implies that a full critical state is not established in the sample at low fields. The field at which the maximum value of $\chi''(\chi')$ tends to saturate depends on the field orientation with respect to the $CuO_2$ plane of YBCO. There is little difference in the peak values of $\chi''(\chi')$ for $H_{rms}$ of 3 Oe and 10 Oe, when H∥c, but in case of H∥ab the difference is significant (Figs 8a-b and Figs 8e-f) where saturation is achieved for $H_{rms} \geq 10$ Oe. The frequency of AC field has little influence on the shape of Cole-Cole plot for a given magnitude of the magnetic field when full critical state has been established.

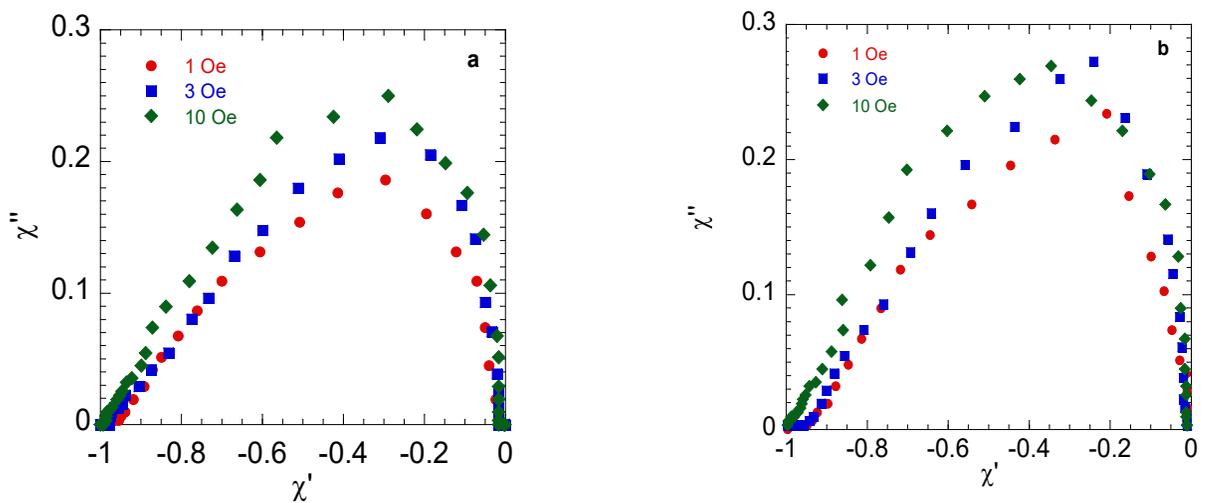

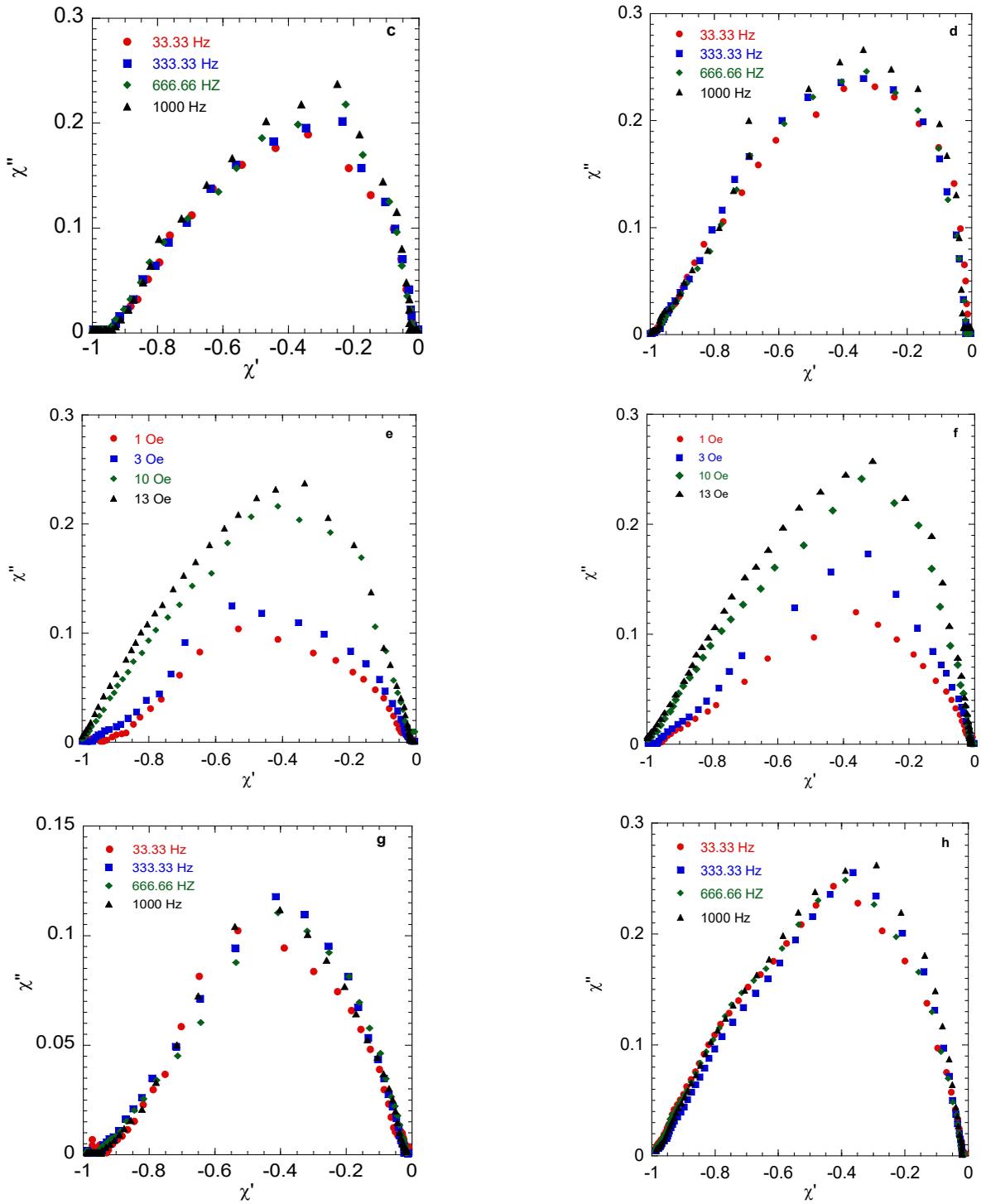

Fig. 8: Field and frequency dependent Cole-Cole plots. (a) H||c (f = 33.33 Hz), (b) H||c (f = 1000 Hz), (c) H||c ($H_{rms}$ = 1 Oe), (d) H||c ($H_{rms}$ = 10 Oe), (e) H||ab (f = 33.33 Hz), (f) H||ab (f = 1000 Hz), (g) H||ab ($H_{rms}$ = 1 Oe) and (h) H||ab ($H_{rms}$ = 13 Oe).

To investigate the effect of the orientation of the magnetic field with respect to the $CuO_2$ plane of YBCO on the profile of $\chi''(\chi')$ in greater depth, we have shown the Cole-Cole plot in the full critical state for both H||c and H||ab together in Fig. 9. It is important to note in Fig. 9 that even though there is a large difference in the demagnetization factors [34] and shielding current paths in the two different field configurations, the $\chi''(\chi')$ features for H||c and H||ab are essentially identical when critical states are

attained. This result is particularly instructive, implying that for high-quality single crystals of YBCO, the geometry of the single crystal has little influence on the shape of the Cole-Cole plot.

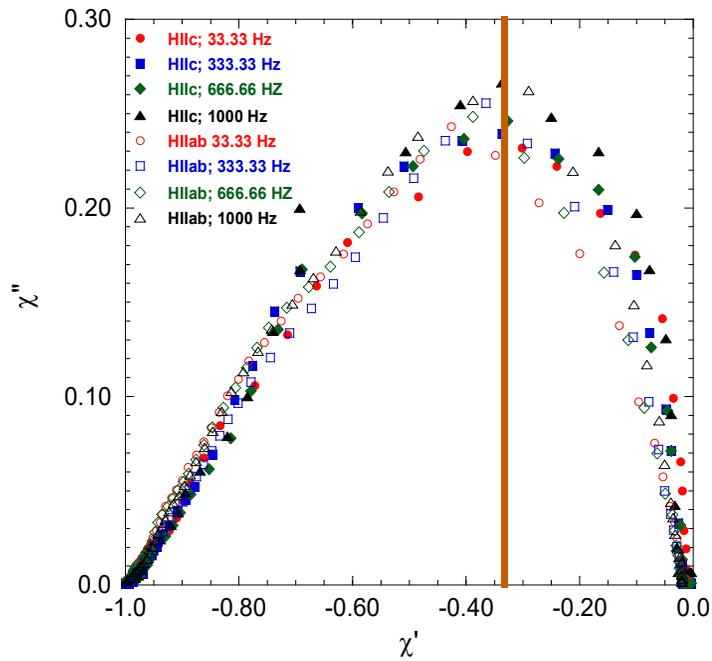

Fig. 9: Cole-Cole plot for both H||c and H||ab field orientations at different AC frequencies. The thick vertical line locates $\chi'$ for the peak value of $\chi''(\chi')$.

## 4. Discussion and conclusions

Fig. 9 illustrates that the peak in $\chi''(\chi')$ for slightly overdoped high-quality YBCO single crystal is located at $\chi' \sim -0.34$ irrespective of the magnetic field orientation when a full critical state has been established. On the other hand, Brandt [20, 21] found the maximum in $\chi''(\chi')$ at $\chi' \sim -0.35$ for axial magnetic fields considering disks and cylinders. Herzog et al. [18] found the maximum in $\chi''(\chi')$ at $\chi' \sim -0.42$ for perpendicular field in case of YBCO thin films. Almost similar result was found by Elabbar [25] for YBCO thin films under perpendicular magnetic fields. Clem and Sanchez [19] found the peak in $\chi''(\chi')$ at $\chi' \sim -0.38$ for perpendicular field considering disk geometry. We summarize these findings in Fig. 10 below. It is instructive to note that Bean CSM predicts the peak in $\chi''(\chi')$ at $\chi' \sim -0.382$ with a maximum value of $\chi'' = 0.241$ [3, 25, 40]. For single crystal under study we have found the maximum value of $\chi''$ within $0.26 \pm 0.01$, close to the value predicted by the Bean CSM.

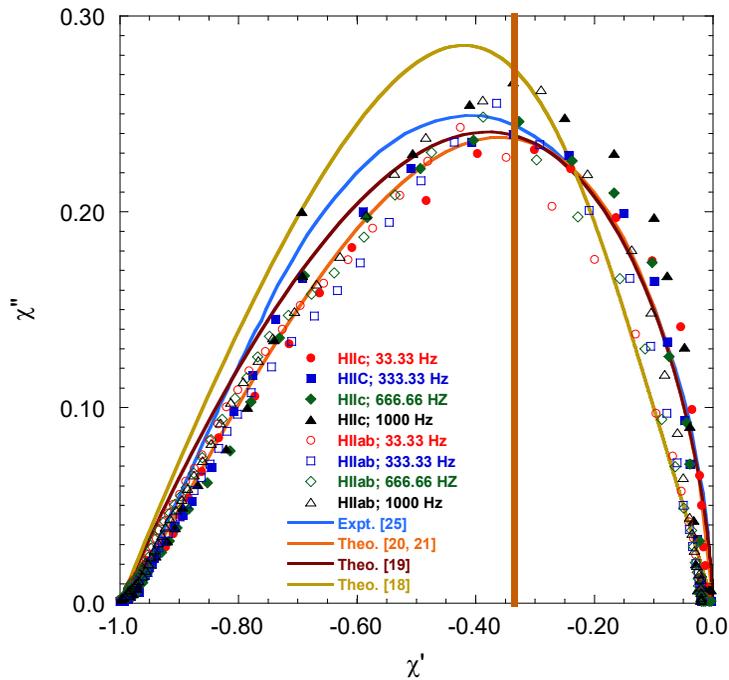

Fig. 10: Cole-Cole plot for both H||c and H||ab field orientations at different AC frequencies for YBCO single crystal together with experimental and theoretical results from earlier studies. The thick vertical line locates $\chi'$ for the peak value of $\chi''(\chi')$.

There have been mainly three types of models suggested in literature in various microscopic physical pictures for the ACS [3, 40]. The first type of model assumes a temperature and field dependent relaxation time $\tau$, that measures how fast a system approaches equilibrium after the disturbance due to the applied AC magnetic field. The resulting expression for complex ACS is of the Debye form [3, 40] and the imaginary part peaks with a value of 0.50 when $\omega\tau = 1$ ($\omega$ is the angular frequency of the AC field). This model assumes fully resistive state and is not suitable for superconductors. The magnitude of the ACS signal and the height of the loss peak are strongly frequency dependent in this scenario. The second type of model emphasizes the diffusive motion of flux lines [3, 40, 46]. When a superconductor shows a linear resistivity, $\rho$, due to diffusive flux motion, regardless of its origins, the penetration depth of the AC field is given by the skin depth, $\delta = (2\rho/\mu_0\omega)^{1/2}$. The maximum magnitude of loss peak in diffusive model is in the range 0.35-0.45 [3, 40, 47] depending on geometry, anisotropy in the resistivity of the sample and the magnetic field orientation. In this model the peak temperature, $T_p$, and magnitude of $\chi''(T)$ change with frequency of the AC field. Furthermore, no field dependence of $T_p$ or magnitude of $\chi''$ is expected in this model. The third model is the non-linear response CSM [3, 40], in which ACS results from the hysteretic penetration of magnetic fluxoids. Hysteretic behavoir is due to the presence of flux pinning centers inside a superconductor. Qualitatively, the peak in $\chi''(T)$ in CSM arises when the magnetic field reaches the center of the sample. $T_p$ is field dependent in CSM but it has no frequency dependence unless $J_c$ is frequency dependent, or there is a possibility of flux creep in the superconductor [3, 40, 43]. The maximum magnitude of the loss peak ratio is ~ 0.24 in CSM when the magnetic field is applied parallel to the surface of a long rectangular slab [3, 25, 40] and $J_c$ is independent of magnetic field H.

Considering the different scenarios described above, our results conform closely to the prediction of the CSM with two moderate deviations. The first one is regarding the position of the peak in the Cole-Cole plot and the second one is related to the peak value of $\chi''$ in the same. As mentioned earlier, in simple CSM with a critical current density independent of the applied magnetic field, the peak value and the peak position of $\chi''$ in the $\chi''(\chi')$ plot are ~0.241 and at $\chi'$ ~-0.382, respectively. For the YBCO single crystal under consideration, the corresponding numbers are ~0.26 and ~-0.34, respectively. A careful modeling of the ACS signal by Shantsev et al. [23] showed that magnetic field dependent $J_c$ can shift the peak in the $\chi''(\chi')$ to the right and enhance the peak value significantly. The model [23] also showed that in the presence of flux creep, where the ACS becomes frequency dependent, the peak value is enhanced again but the shift in the peak position with respect to that in the simple CSM is to the left. Therefore, we believe that the deviations in the Cole-Cole plot for the YBCO single crystal under consideration from the predictions of simple CSM is due to a weak magnetic field dependent critical current density present in this compound.

Close to $T_c$, the flux flow and flux creep phenomena are thermally assisted. The prominence of these phenomena depends strongly on the pinning strength of the superconducting compound. For the slightly overdoped single crystal under consideration, we have not found any appreciable effect of frequency on the ACS. This implies that flux creep is absent here. This is in contrast with other experimental results [17, 19-21, 25-28]. The single crystal under consideration is almost fully oxygen loaded and highly homogeneous (as indicated by the very narrow SC transition width), as a consequence the SC condensation energy is expected to be very high. High SC condensation energy implies high value of the flux pinning potential [2, 45, 48-50] and as a consequence flux creep is strongly inhibited.

The field dependences of $T_p$, $T_d$, and FWHM of $d\chi'(T)/dT$ (Figs. 5 and 6) disclose that the effect of magnetic field is relatively stronger for H∥c configuration compared to that for H∥ab. This indicates that the flux dynamics is anisotropic and flux pinning is comparatively weaker when the magnetic field is applied along the c-axis. The correspondence between $T_p$ and $T_d$ (Fig. 7) is interesting and the phenomenon is possibly related to the formation of the critical state and propagation of the flux front within the sample as temperature is lowered [51].

To conclude, we have presented a detailed analysis of the ACS of high-quality slightly overdoped YBCO single crystal with different orientations of the magnetic field. No appreciable sign of flux creep has been found for the magnetic field and frequency ranges considered. The $\chi''(\chi')$ features are consistent with the CSM with a weakly field dependent $J_c$.

**Acknowledgements**

The experimental work has been carried out in the International Research Centre in Superconductivity (IRCS) and S.H.N. thanks the IRCS, University of Cambridge, UK for providing with the facilities. S.H.N. thanks Professor J.R. Cooper for his inspirational inputs. S.H.N. also acknowledges the research grant (1151/5/52/RU/Science-07/19-20) from the Faculty of Science, University of Rajshahi, Bangladesh, which supported the theoretical part of this work.

**Data availability**

The data sets generated and/or analyzed in this study are available from the corresponding author on reasonable request.

**List of references**

[1] A. Semwal, N.M. Strickland, A. Bubendorfer, S.H. Naqib, S.K. Goh, G.V.M. Williams, Supercond. Sci.Technol. 17 (2004) S506.

[2] S.H. Naqib, A. Semwal, Physica C 425 (2005) 14.

[3] F. Gomory, Supercond. Sci. Technol. 10 (1997) 523.

[4] M. Golosovsky, M. Tsindlekht, D. Davidov, Supercond. Sci. Technol. 9 (1996) 1.

[5] M.D. Ainslie, H. Fujishiro, Supercond. Sci. Technol. 28 (2015) 053002.

[6] Ruslan Prozorov, Russell W. Giannetta, Supercond. Sci. Technol. 19 (2006) R41.

[7] C.P. Bean, Rev. Mod. Phys. 2 (1964) 31.

[8] P.W. Anderson, Y.B. Kim, Rev. Mod. Phys. 36 (1964) 39.

[9] K.H. Muller, Physica C 159 (1989) 717.

[10] J.R. Clem, Physica C 50 (1988) 153.

[11] L. Ji, R.H. Sohn, G.C. Spalding, C.J. Lobb, M. Tinkham, Phys. Rev. B 40 (1989) 10936.

[12] T. Ishida, R.B. Goldfarb, Phys. Rev. B 41 (1990) 8937.

[13] C.Y. Lee, Y.H. Kao, Physica C 241 (1995) 167.

[14] G. Blatter, M.Y. Feigel'man, Y.B. Geshkenbein, A.I. Larkin, V.M. Vinokur, Rev. Mod. Phys. 66 (1994) 1125.

[15] Wai-Kwong Kwok, Ulrich Welp, Andreas Glatz, Alexei E Koshelev, Karen J. Kihlstrom, George W. Crabtree, Rep. Prog. Phys. 79 (2016) 116501.

[16] E.H. Brandt, Rep. Prog. Phys. 58 (1995) 1465.

[17] O. Stoppard, D. Gugan, Physica C 241 (1995) 375.

[18] Th. Herzog, H.A. Radovan, P. Ziemann, E.H. Brandt, Phys. Rev. B 56 (1997) 2871.

[19] J.R. Clem, A. Sanchez, Phys. Rev. B 50 (1994) 9355.

[20] E.H. Brandt, Phys. Rev. B 58 (1998) 6523.

[21] E.H. Brandt, Phys. Rev. B 58 (1998) 6506.

[22] B.J. Jonsson, K.V. Rao, S.H. Yun, U.O. Karlsson, Phys. Rev. B 58 (1998) 5862.

[23] D.V. Shantsev, Y.M. Galperin, T.H. Johansen, Phys. Rev. B 61 (2000) 9699.

[24] Y.Q. Qi, L.L. Liu, J. Supercond. Nov. Magn 28 (2015) 1749.

[25] A.A. Elabbar, Physica C 469 (2009) 147.

[26] G.C. Han, C.K. Ong, S.Y. Xu, H.P. Li, Appl. Phys. Lett. 71 (1997) 1860.

[27] E.K. Nazarova, A.J. Zaleski, K.A. Nenkov, A.L. Zahariev, Physica C 468 (2008) 955.

[28] M.G. das Virgens, S. García, M.A. Continentino, L. Ghivelder, Phys. Rev. B 71 (2005) 064520.

[29] S.H. Naqib, J.R. Cooper, J.L. Tallon, C. Panagopoulos, Physica C 387 (2003) 365.

[30] S.H. Naqib, J.R. Cooper, J.L. Tallon, R.S. Islam, R.A. Chakalov, Phys. Rev. B 71 054502.

[31] S.H. Naqib, R.S. Islam, Supercond. Sci. Technol. 21 (2008) 105017.

[32] C. Changkang, H. Yongle, J.W. Hodby, B.M. Wanklyn, A.V. Narlikar, S.B. Samanta, J. Mater. Sci. Lett. 15 (1996) 886.

[33] D. Babić, J.R. Cooper, J.W. Hodby, Chen Changkang, Phys. Rev. B 60 (1999) 698.

[34] S.H. Naqib, C.P.G.S. thesis (1999) University of Cambridge (unpublished).

[35] S.D. Obertelli, J.R. Cooper, J.L. Tallon, Phys. Rev. B 46 (1992) 14928.

[36] M.R. Presland, J.L. Tallon, R.G. Buckley, R.S. Liu, N.E. Flower, Physica C 176 (1991) 95.

[37] W. Xing, B. Heinrich, J. Chrzanowski, J.C. Irwin, H. Zhou, A. Cragg, A.A. Fife, Physica C 205 (1993) 311.

[38] V.A. Atsarkin, G.A. Vasneva, S.G. Zybtsev, Supercond.Sci.Technol. 7 (1994) 423.

[39] M. Daumling, D.C. Latbalesteir, Phys. Rev. B 40 (1990) 9350.

[40] *Magnetic Susceptibility of Superconductors and Other Spin systems*, Eds. R.A. Hein, T.L. Francavilla, D.H. Liebenberg (New York, Plenum 1991).

[41] C.P. Bean, Phys. Rev. Lett. 8 (1962) 250.

[42] H. London, Phys. Lett. 6 (1963) 162.

[43] K.H. Muller, Physica C 168 (1990) 585.

[44] M.R.H. Sarkar, S.H. Naqib, J. Sci. Res. 4 (2) (2012) 287.

[45] S.H. Naqib, R.S. Islam, Chin. Phys. B 24 (2015) 017402.

[46] E.H. Brandt, Z. Phys. B 80 (1990) 167.

[47] F. Supple, A.M. Campbell, J.R. Cooper, Physica C 242 (1995) 233.

[48] S.H. Naqib, R.S. Islam, Scientific Reports 9 (2019) 14856.

[49] J.L. Tallon, G.V.M. Williams, J.W. Loram, Physica C 338 (2000) 9.

[50] Nazir Ahmad, S.H. Naqib, Results in Physics 17 (2020) 103054.

[51] J.Z. Sun, M.J. Scharen, L.C. Bourne, J.R. Schrieffer, Phys. Rev. B 44 (1991) 5275.

**Author Contributions**

M.R.H.S. performed the data analysis and contributed in draft writing. S.H.N. designed and supervised the project, measured the ACS, and finalized the manuscript. Both the authors reviewed the manuscript.

**Additional Information**

**Competing Interests**

The authors declare that they have no known competing financial interests or personal relationships that could have appeared to influence the work reported in this paper.